\documentstyle[12pt]{article}

\newcommand{\bt}{\beta}
\newcommand{\beg}{\begin{equation}}
\newcommand{\enq}{\end{equation}}
\newcommand{\sig}{\sigma}
\newcommand{\cst}{{\rm constant}}

\begin{document}

\title{Asymptotic Level Spacing Of The Laguerre Ensemble: A Coulomb Fluid
Approach}
\author{Y. Chen and S. M. Manning\\
        Department of Mathematics\\
         Imperial College\\
         London SW7 2BZ\\
              U. K. }
\date{\today}
\maketitle

\begin{abstract}
We determine the asymptotic level spacing distribution for the
Laguerre Ensemble in a single scaled interval,
$(0,s)$, containing no levels,
$E_{\bt}(0,s)$, via Dyson's Coulomb Fluid approach. For the $\alpha=0$
Unitary-Laguerre Ensemble, we recover the exact spacing distribution found
by both Edelman and Forrester, while for $\alpha\neq 0$, the leading
terms of $E_{2}(0,s)$, found by Tracy and Widom,
are reproduced without the use of the Bessel kernel and the associated
Painlev\'e transcendent. In the same
approximation, the next leading term, due to a ``finite temperature''
perturbation $(\bt\neq 2)$, is found.
\end{abstract}

\newpage

\noindent
{\bf Introduction}
\par\noindent
The probability that there are no levels in a scaled interval $(-t, t)$
[where $t$ is measured
with respect to the averaged spacing], $E(0,t)$,
in a long stack of energy levels of heavy nuclei was given by a
conjecture of Wigner\cite{ref:Wigner} and was well
supported by experimental data\cite{ref:Porter} for systems with time
reversal symmetry. In a series of seminal papers, Dyson\cite{ref:Dyson}
introduced a new class of random matrix ensembles and determined in the
continuum approximation
(expected to be valid when number of levels, $N$, is very large) that
$\ln E_{\bt}(0,t)\sim -\frac{\pi^2}{4}\bt t^2-(1-\bt/2)\pi t$, using
the methods of classical electrostatics, potential theory and
thermodynamics, for ensembles with unitary $(\bt=2)$, orthogonal
$(\bt=1)$ and symplectic $(\bt=4)$ symmetries.
A term of $O(\ln t)$ and a constant missed in
the continuum approximation was later discovered by
Widom, des Cloizeaux and Mehta, and Dyson
\cite{ref:Widom,ref:des Cloizeaux,ref:Dyson76,ref:remark,ref:comment}.
\par\noindent
Recently, in a remarkable paper, Tracy and Widom\cite{ref:Tracy}
showed, in the single interval case, that the logarithmic derivative
of the Fredholm determinant of
the Bessel Kernel---which arises in the scaling limit [with respect to
the hard edge density, see\cite{ref:Tracy,ref:remark1}] of the unitary
Laguerre ensemble---satisfies a Painlev\'e V equation, from which
the asymptotic level spacing distribution can be computed exactly,
amongst other quantities of interest from the random matrix point of
view\cite{ref:footnote}.
\par\noindent
In this letter we shall employ the continuum approximation of Dyson to
calculate the level spacing distribution for the Laguerre ensemble
[with $\alpha \neq 0$, see\cite{ref:Tracy}], with $\bt = 2$.
For $\bt\neq 2$, the spacing distribution can be found by a
perturbative calculation due to Dyson\cite{ref:Dyson}.
\par\noindent
{}From the Brownian motion model, in the ``hydrodynamical''
approximation, Dyson\cite{ref:Dyson1} derived an equation satisfied
by the non-equilibrium level density $\sig(x,\tau)$;
\beg
{\partial\over \partial\tau}\sig(x,\tau)={\partial\over \partial x}
\left(\bt \sig(x,\tau){\partial \Psi\over \partial x}\right),
\quad\quad \tau >0
\enq
with
\beg
\Psi(x,\tau)=\frac{1}{\bt}u(x)-\int dy\sig(y,\tau)\ln|x-y|+
\left(\frac{1}{\bt}-\frac{1}{2}\right)\ln\left[\sig(x,\tau)\right],
\enq
where the fictitious time $\tau$ pulls the levels towards the observed
level density, $r(x)$, generated by the imposed potential
$u(x)/\bt=\int dy r(x)\ln |x-y|$, (as $\tau\rightarrow \infty$),
that holds the Coulomb Fluid together.
The stationary solution, reached as $\tau\rightarrow \infty$, for the
level density statisfies the H\"ukel-like self-consistent equation
\beg
u(x)-\bt\int dy \sig(y)\ln |x-y|+\left(1-\frac{\bt}{2}\right)
\ln\left[\sig(x)\right]=A=\cst,
\label{eq:sig}
\enq
with effective temperatute $T_{e} \propto (1-\frac{\bt}{2})$.
This, in turn, maybe derived from the following variational
principle: $\min_{\sig,\mu}F[\sig,\mu],$ with
\beg
F[\sig, \mu]=\bt \int dx\Psi(x)\sig(x)-\mu\left(\int
dx\sig(x)-N\right),
\enq
and $A=-\mu+(1-\bt/2),$ where $\mu$ is the chemical potential.
Therefore the free energy, $F$, at equilibrium, with exactly $N$ levels
contained in an interval $I$ is
\beg
F[I]=\frac{1}{2}AN+\frac{1}{2}\int_{I}dx\;u(x)\sig(x)-\frac{1}{2}
\left(1-\frac{\bt}{2}\right)\int_{I}dx\sig(x)\ln[\sig(x)],
\label{eq:F_def}
\enq
subject to $\int_{I}dx\sig(x)=N.$
For the Laguerre Ensemble, $u(x)=x-\alpha \ln x,$
$x\in (0,\infty)$, but as we are using the continuum
approximation, an upper band edge, $b \in (0,\infty)$
 must be imposed on the
level density to produce a finite number of levels. In the large $N$
limit, the level density is $\sig(x)=\frac{1}{4\pi}{\sqrt {{4N-x}\over
x}}$, $x\in (0,4N)$ for the unitary case\cite{ref:Bronk,ref:Nagao}.
This is distinct from the Wigner semi-circle law, where $u(x)=x^2$,
$x\in(-\infty,+\infty)$.
\par\noindent
{\bf Asymptotic Spacing }
\par\noindent
The probability distribution that an interval $(0,a)$
contains no levels is, by definition, the ratio of the partition
function where all $N$ levels reside in the complement of $(0,a)$
i.e. $(a,b)$ [where $b$ is the upper band edge and $0<a<b$] to that
for which all $N$ levels reside in the full interval i.e. $(0,b)$,
and is\cite{ref:Dyson},
\beg
\ln E_{\bt}(0,a)=-[F(a,b)-F(0,b)].
\enq
In the continuum
approximation, $F$ is given by Eq.(\ref{eq:F_def}), where $\sig(x)$ solves
Eq. (\ref{eq:sig}) \cite{ref:remark3}.
{}From thermodynamic considerations, since
$F(a,b)$ is the Free Energy in a ``constricted'' region, we must have
$F(a,b)>F(0,b).$ We now proceed
to solve Eq. (\ref{eq:sig}) for $\bt=2$, in the interval $(a,b)$.
To simplify the mathematics, Eq.(\ref{eq:sig}) is converted into
a singular intergral equation by taking a derivative with respect to
$x$, and can be solved by a standard method\cite{ref:singular}.
The solution of this equation is
\begin{equation}
\sig(x) = \frac{1}{\pi^2\bt}{\sqrt {{b-x\over x-a}}}\int_{a}^{b}
\frac{dy}{y-x}{\sqrt {{y-a\over b-y}}}\left(1-\frac{\alpha}{y}\right)
\label{eq:sig1}
=\frac{1}{\pi\bt}{\sqrt {{b-x\over x-a}}}\left(1-\frac{\alpha}{x}
{\sqrt {{a\over b}}}\right)\;,
\label{eq:sig_soln}
\end{equation}
where $x\in(a,b)$ and to maintain positivity we demand that
$\sqrt{ab} > \alpha.$ A
straightforward calculation supplies the normalization condition:
\beg
N=\frac{b-a}{2\bt}+\frac{\alpha}{\bt}\left[{\sqrt {{a\over b}}}-1\right].
\label{eq:N}
\enq
We have deliberately left $\bt$ without setting it equal to $2$ in
Eqs. (\ref{eq:sig_soln}) and (\ref{eq:N}). It is clear from the
structure of Eq.(\ref{eq:N}), that $N$, the total number of levels, is almost
exhausted by the first term, but
as to be seen later, the second term cannot be discarded.
As we are required to compare
$F(a,b)$ with $F(0,b)$, where $N$ is very large, to
facilitate the computation we shall evaluate instead, $F(a,b)$
in the limits $a\ll b,\;\frac{a}{b-a}\ll 1\; {\rm
and}\;\frac{b}{b-a}\sim 0(1)$, thus by-passing an independent
computation of $F(0,b)$.
To evaluate $F(a,b)$ requires the
determination of $A$ and the ``interaction energy''
$\frac{1}{2}\int_{a}^{b}dx\sig(x)u(x)$. In the limits stated we shall extract
the {\it very large} terms, which are functions of $N$ only and finite
terms that are functions of $Na$ only; any remaining terms are
therefore negligable in the large $N$ limit. Note that the very large
terms are then subtracted according to the definition of
$E_{\bt}(0,a)$, leaving behind terms only those terms in $Na$.
\par\noindent
For the constant $A$, we
send $s\rightarrow b^-$ in Eq.(\ref{eq:sig}) [with $\bt=2$], which gives
\beg
A=a-\frac{b-a}{2}\ln \left[\frac{b-a}{4e}\right]
-\alpha \ln b-\alpha\left[{\sqrt {{a\over b}}}-1\right]\ln (b-a)+
\frac{\alpha}{\pi}{\sqrt {a\over b}}I\left(\frac{a}{b-a}\right),
\enq
where
\begin{eqnarray}
I(x) &:= & \int_{0}^{1}dt\;{\sqrt {1-t\over t}}\;{\ln (1-t)\over t+x}
	= {1\over x}{\partial\over \partial \bt}
\left[B\left(\frac{1}{2},\bt\right)
F\left(1,\frac{1}{2}, \frac{1}{2}+\bt,-\frac{1}{x}\right)
\right]\bigg |_{\bt =3/2} \nonumber \\
     & = & \sim  -2\pi(1-\ln 2)+\pi\sqrt{x}+\cdots,\;x\ll 1.
\end{eqnarray}
\par\noindent
As $N\rightarrow \infty$, but with $Na$ finite,
we find that
\beg
\frac{1}{2}NA\sim -N^2\ln (N/e)-\frac{\alpha}{2}\ln (16N/e^4)
+\frac{1}{2}Na-\frac{\alpha}{2}{\sqrt {Na}}\;,
\label{eq:NA}
\enq
where to capture all the terms in $Na$ it is essential to use
Eq.(\ref{eq:N}) for $N$.
For the contribution due to the interaction, we have (for $\bt=2$)
\begin{eqnarray}
\lefteqn{\frac{1}{2}\int_{a}^{b}dt\;\sig(t)\;u(t) =
\frac{1}{2}\left[{(b-a)\over 2\bt}\right]^2+
\frac{1}{2}\,a\,{(b-a)\over 2\bt}
-{\alpha\over 2}\frac{(b-a)}{2\bt}\left[{\sqrt {{a\over b}}}-1\right]
} \nonumber \\
& &
-\frac{\alpha}{\pi}{(b-a)\over 2\bt}\;h\left({a\over b-a}\right)
-{\alpha^2\over 2\bt}\ln (b-a)\left[{\sqrt {{a\over b}}}-1\right]
+{\alpha^2\over 2\bt\pi}{\sqrt {{a\over b}}}\;g\left({a\over b-a}\right),
\end{eqnarray}
where
\begin{equation}
h(x) := \int_{0}^{1}dt\;{\sqrt {{1-t\over t}}}\;\ln (t+x)
     \sim -\frac{\pi}{2}\ln (4e)-\pi\;x +2\pi {\sqrt x}+\cdots
,\quad\quad x\ll 1
\end{equation}
and
\begin{eqnarray}
g(x) & := & \int_{0}^{1}{dt\over t+x}\;{\sqrt {{1-t\over t}}}\ln (t+x)
	\nonumber \\
     	& = & -\frac{\pi}{2}\frac{\partial}{\partial \rho}
	\left[x^{-\rho}
	F\left(\rho,\frac{1}{2};2;-1/x\right)\right]\bigg|_{\rho=1}
	 \sim \frac{\pi}{\sqrt{x}}\ln (4x) + \pi\ln x + \cdots
,\; x\ll 1
\end{eqnarray}
With the same considerations,
\beg
\frac{1}{2}\int_{a}^{b}dt\;\sig(t)\; u(t)
\sim
\frac{1}{2}N^2-\alpha N\ln (2/e^{1/2})-\frac{\alpha^2}{2}\ln 4N+
\frac{1}{2}Na-\frac{3}{2}\alpha{\sqrt {Na}}+\frac{\alpha^2}{4}\ln (4Na).
\label{eq:interaction}
\enq
Pooling together Eqs.(\ref{eq:NA}) and (\ref{eq:interaction}) we find
by subtracting the very large terms
\beg
F(a,b)-F(a\ll b,b)
 \sim Na-2\alpha {\sqrt {Na}}+\frac{\alpha^2}{4}\ln(4Na),
\enq
which gives,
\beg
E_{2}(0,s)\sim {{\rm e}^{-s/4+\alpha {\sqrt s}}\over s^{\alpha^2/4}},
\label{eq:spac_0T}
\enq
upon scaling with respect to the hard edge density, i.e. with the
replacement $Na\rightarrow \frac{s}{4}.$
\par\noindent
Before we proceed to give the result for the $\bt\neq 2$ case, we should
like to mention that by repeating the calculation in the far simpler
situation where $\alpha =0$, Eq.(\ref{eq:N}) can be solved trivially, which
gives $b-a=4N$ [for $\bt=2$] and the change in the
free energy is equal to $Na$ or $\frac{s}{4}$. This is the exact result of
Edelman\cite{ref:Edelman} and of
Forrester\cite{ref:Forrester}; $E_2(0,s)={\rm e}^{-s/4},\;\alpha=0,\;
s\in (0,\infty).$ For $\alpha\neq 0$, $\bt=2$,
Tracy and Widom\cite{ref:Tracy}, found through an asymptotic expansion
of a Painlev\'e V equation, higher order terms, in $1/{\sqrt s}$, $1/s$
etc., and made a conjecture concerning the term independent of $s$.
It was observed by Tracy and Widom that\cite{ref:Tracy} with
$\alpha=\mp\frac{1}{2}$, the Bessel kernel reduces to the
kernels which arise by scaling into the bulk of the spectrum of the
Gaussian orthogonal and symplectic ensembles, respectively, provided
one makes the replacement $s\rightarrow \pi^2 t^2$.
It is amusing to see that
by approching the GOE and GUE through the ``back door'', i.e. using
the mapping of Tracy and Widom\cite{ref:Tracy}; the Coulomb
Fluid approach, when applied to the Laguerre ensemble, supplies rather
precise information\cite{ref:remark4}.
\par\noindent
{\bf Correction to the Free Energy When $\bt\neq 2$}
\par\noindent
This is simply found by adding the finite temperature contribution
\beg
\frac{\delta F(a,b)}{\left(1-\frac{\bt}{2}\right)}
=\int_{a}^{b}dx\sig(x)\ln
[\sig(x)]
\sim -\ln \left[\frac{2\pi}{e}\right]N-\frac{\alpha}{2}\ln
4N+\frac{\alpha}{4}\ln 4Na,
\quad\quad a\ll b-a,
\enq
to the $\bt=2$ Free Energy.
\par\noindent
Collecting the appropriate terms together, we find that
\beg
-\ln E_{\alpha}(0,s)\sim \frac{s}{2\bt}-\frac{2\alpha}{\bt}{\sqrt s}
+\frac{\alpha^2}{2\bt}\ln s
+\left(1-\frac{\bt}{2}\right)\frac{\alpha}{4}\ln s,
\enq
which clearly reduces to Eq.(\ref{eq:spac_0T}) when $\bt=2.$
\par\noindent
{\bf Conclusion}
\par\noindent
We should like to mention that by treating the constraint, i.e.
Eq.(\ref{eq:N}) more accurately, it should be possible to produce
the higher order correction terms $1/{\sqrt s},\;1/s\ldots $ mentioned
previously. However, the computation would become exceedingly
complicated and clearly the methods of Tracy and
Widom\cite{ref:Tracy} have much to be desired.
\par\noindent
The above calculations suggest upon comparison with exact results
that the Coulomb Fluid approach is quite robust and may shed light on the
level spacing distribution of the $q-$Laguerre ensemble, which arises
in the context of electronic transport in disordered systems
\cite{ref:remark2}, with the potential
\beg
u(x,q)=\sum_{n=0}^{\infty}\ln\left[1+(1-q)x\;q^n\right],\;\;\; q\in(0,1)
\enq
which reduces to the ordinary Laguerre potential as $q\rightarrow
1^-$; $u(x,1^-)=x.$
We leave as a future project the determination of
the level spacing distribution with $u(x;q).$

\vfill\eject


\begin{thebibliography}{99}
\bibitem{ref:comment} As will be seen later, the computation of
$E_{\bt}(0,t),$ involves subtraction of two large terms which are
functions of $N$ and the continuum approximation is not expected to
have sufficiently fine resolution to determine the constant.
\bibitem{ref:remark3}For a discussion on the continuum approximation,
the reader is urged to consult the second paper of \cite{ref:Dyson}.
\end{thebibliography}
\end{document}